\pdfoutput=1
\RequirePackage{amsmath} 
\documentclass{llncs}
\usepackage[utf8]{inputenc}
\usepackage{mathtools}
\usepackage{multirow}
\usepackage{xfrac}
\usepackage{nicefrac}
\usepackage{pbox}
\usepackage[normalem]{ulem} 
\usepackage{color}
\usepackage{graphicx}
\graphicspath{{./images/}}
\usepackage[acronym,nomain]{glossaries}
\usepackage[pdftex,
            pdfusetitle,
            pdfauthor={Axel Huebl, Rene Widera, Felix Schmitt, Alexander Matthes, Norbert Podhorszki, Jong Youl Choi, Scott Klasky, Michael Bussmann},
            pdfsubject={Scientific Paper in High-Performance Computing and Modeling},
            pdfkeywords={HPC IO Scalability Data Reduction ADIOS HDF5 Compression Staging GPU PIConGPU Manycore Exascale},
            pdfproducer={Latex with hyperref},
            pdfcreator={pdflatex}]{hyperref}
\newcounter{numrel}

\newacronym{pic}{PIC}{particle-in-cell}
\newacronym{ornl}{ORNL}{Oak Ridge National Laboratory}

\begin{document}

\title{On the Scalability of Data Reduction Techniques in Current and Upcoming HPC Systems from an Application Perspective\thanks{
  This project has received funding from the European Unions Horizon 2020 research and innovation programme under grant agreement No 654220.
  An award of computer time was provided by the Innovative and Novel Computational Impact on Theory and Experiment (INCITE) program.
  This research used resources of the Oak Ridge Leadership Computing Facility at the Oak Ridge National Laboratory, which is supported by the Office of Science of the U.S. Department of Energy under Contract No. DE-AC05-00OR22725.
}}

\author{Axel Huebl\inst{1,2}(0000-0003-1943-7141)
    \and Ren{\'e} Widera\inst{1}(0000-0003-1642-0459)
    \and Felix Schmitt\inst{2,3}
    \and Alexander Matthes\inst{1,2}(0000-0002-6702-2015)
    \and Norbert~Podhorszki\inst{4}(0000-0001-9647-542X)
    \and Jong Youl Choi\inst{4}
    \and Scott Klasky\inst{4}
    \and Michael Bussmann\inst{1}(0000-0002-8258-3881)
}
\institute{Helmholtz-Zentrum Dresden -- Rossendorf, Germany\\
    \email{\{a.huebl,m.bussmann\}@hzdr.de}
	\and Technische Universit\"at Dresden, Germany
	\and NVIDIA ARC GmbH, Germany
    \and Oak Ridge National Laboratory, United States of America
}
\maketitle
\begin{abstract}
We implement and benchmark parallel I/O methods for the fully-manycore driven particle-in-cell code PIConGPU.
Identifying \sloppy{throughput} and overall I/O size as a major challenge for applications on today's and future HPC systems, we present a scaling law characterizing performance bottlenecks in state-of-the-art approaches for data reduction.
Consequently, we propose, implement and verify multi-threaded data-transformations for the I/O library ADIOS as a feasible way to trade underutilized host-side compute potential on heterogeneous systems for reduced I/O latency.
\end{abstract}

\section{Introduction}

Production-scale research simulation codes have been optimized in the last years to achieve maximum compute performance on leadership, heterogeneous computing systems such as the Titan supercomputer at \gls{ornl}.
With close to perfect weak scaling domain scientists can increase spatial and temporal resolution of their simulation and explore systems without reducing dimensionality or feature resolution.

We present the consequences of near-perfect weak-scaling of such a code in terms of I/O demands from an application perspective based on production runs using the \gls{pic} code PIConGPU~\cite{PIConGPU2013,PIConGPUlatest}.
PIConGPU demonstrates a typical use case in which a PFlops/s-scale, performance portable simulation~\cite{zenker2016alpaka,OpenPower2016} leads automatically to PByte-scale output even for single runs.


\subsection{PIConGPU}

PIConGPU is an electro-magnetic \gls{pic} code~\cite{birdsall1985,hockney1988} implemented via abstract, performance portable C++11 kernels on manycore hardware utilizing the Alpaka library~\cite{zenker2016alpaka,OpenPower2016}.
Its applications span from general plasma physics, over laser-matter interaction to laser-plasma based particle accelerator research.

Since its initial open-source release in 2013 with CUDA support, PIConGPU is reportedly the fastest particle-in-cell code in the world in terms of sustained peak Flops/s~\cite{PIConGPU2013}.
We achieved this by not only porting the bottlenecks of the \gls{pic} algorithm to new compute hardware but the complete code, thus minimizing data transfer.
PIConGPU data structures are tiled and swapping of frequently updated data residing on device memory over low-bandwidth bottlenecks such as the PCI bus is avoided~\cite{burau2010}.


The overall simulation is spatially domain decomposed and only nearby border areas need to be communicated across compute nodes (and accelerators) inbetween iterations.
Iterations in PIConGPU are performed with a frequency of about 10\,Hz on current accelerator architectures (GPUs) when simulating 3D spatial domains and up to 60\,Hz for two-dimensional domains.
Each iteration updates electro-magnetic fields and plasma particles, which together constitute the simulation's state.

\subsection{Physical Observables}

We will define primary observables as variables directly accessible and iterated within the simulation.
In terms of an electro-magnetic \gls{pic} code these are electric field, magnetic field and plasma particles' properties such as position, momentum, charge to mass ratio and weighting.
Primary observables are convenient for the domain expert for exploration, of limited use for theories and models and nearly always inaccessible directly in experiments.

We define secondary observables as computable on-the-fly, \gls{pic} examples being the electric current density, position-filtered energy histograms or projected phase space distributions.
In practice, analysis of a specific setup needs multiple additional, study-specific derivations from already derived observables which we summarize as tertiary observables.
Examples in the domain of plasma physics are integrals over phase-space trajectories, time-averaged fields, sample trajectories or particle distributions in gradients of fields, flux over time, growth-rates, etc..
Usually, observables accessible by experiments fall in this last category and can be compared to theoretical model predictions.

\subsection{Two Example Workflows to Explore Complex Systems}

In daily modeling work we usually iterate between two operational modes while investigating a new physical system.
We start with an exploratory phase guided by initial hypotheses, looking at primary observables via visualizations or utilizing existing analysis pipelines to iterate over the result of strongly reduced secondary and tertiary observables.
During this phase, we develop new study-specific analysis steps and working hypotheses.

The second phase continues with a high-resolution, high-throughput scan of an identified regime of the physical system to prove or falsify our working hypotheses.
Due to higher resolution and full physical modeling, new observations will emerge from that step.
Research is then about iterating both steps in a refined manner until a system is well understood and a model is found to describe the complex processes of interest.

\subsection{Structure of this Paper}

As our guiding example, we describe the Titan and Summit systems at \gls{ornl} and their I/O bandwidth hierarchies from the special perspective of a fully GPU-driven, massively parallel \gls{pic} code.
We then evaluate the performance of PIConGPU's I/O implementation, the overhead it introduces and mitigation strategies via on-the-fly data reductions.
We address issues in current state-of-the-art compression schemes for our application and compare them to self-implemented compression schemes that make optimal use of underutilized hardware components.
More specifically, we integrated the meta compression library blosc~\cite{blosc} into ADIOS, thereby for the first time enabling multi-threaded compression within ADIOS.

\section{ORNL Titan \& Summit Systems}

With the launch of the Titan supercomputer to the public in 2013, manycore powered supercomputing finally became accessible on large-scale installations.
Since then, the share of accelerator hardware in the TOP\,100 systems has risen to one third~\cite{TOP500}.
Such heterogeneous systems concentrate their compute performance in the accelerator component, usually outnumbering the host system's compute potential by an order of magnitude, a trend that seems to continue on upcoming systems such as Summit.

\subsection{I/O Limitations in State-of-the-Art Systems}
\label{sec:iobottlenecks}

The parallel file system Atlas at \gls{ornl}, partitioned in two islands of 14 PBytes each, provides an overall design parallel bandwidth of $B_\text{parallel}=1$~TByte/s.
It is worth noting that \textit{if} a hypothetical application would be constantly writing at this maximum parallel bandwidth, Atlas would run out of disk memory in less than 9\,hours.
We managed to write within each 8\,hour production run of our plasma simulation code PIConGPU about 1 PByte of (zlib) compressed data, sampling the full system state every 2000 iteration steps. PIConGPU thus presents a realistic use case that can consume a significant fraction of those resources.
With the upper limit of shared storage in mind, it is clear that data reduction comes with great value.
Additionally, fast migration to and from tape storage and a strictly imposed short data lifetime on Atlas also encourage users to avoid occupying disk memory for too long.

An equally severe limitation for I/O besides maximum data size is the overall time $t_\text{I/O}$ for file I/O compared to one iteration of the simulation, including data preparation time $t_\text{prep}$.
Compared to the time $t_\text{without\;I/O}$ one iteration takes without I/O this $t_\text{I/O}$ introduces an overhead to the application run time, so that the single iteration runtime with I/O becomes $t_\text{with\;I/O}=t_\text{without\;IO}+t_\text{I/O}$.
When considering applications scaling to the full Titan system, reaching TByte/s overall throughput results in a maximum node-average throughput of 55\,\allowbreak{}MByte/s.
Applications with near perfect scaling can generate GPU data at two-digit Hz levels amounting to data rates as high as \sloppy{$10\times 6$\,GiByte/s} (device global memory) on a node-local level, outnumbering the file system performance by three orders of magnitude.
Asynchronous I/O lowers this dramatic gap temporarily, but still throttles the application at least to 1/10th of the bandwidth of the CPU-GPU interconnect, not accounting for data reorganization from tiled GPU memory to per-node contiguous memory as expected by parallel I/O APIs.

\subsection{Staging, Burst Buffers and I/O Backlog}
\label{sec:backlog}

Even at moderate data rates, asynchronous writing can quickly overlap with the next consecutive write period.
Staging\cite{DataStager,DataSpaces}, if operating off-node, can reduce that data pressure but is similarly limited by another order of magnitude gap in throughput as soon as the interconnect is accessed.

Systems such as NERSC's Cori recently introduced so-called burst-buffers~\cite{NERSCburstBuffers}.
Located either off-node similar to I/O nodes or in-node as with the upcoming Summit system, overall size of those burst buffers is usually similar to that of the global host RAM with access bandwidth ranging between network-interconnect and parallel filesystem bandwidth.

Burst buffers provide an interesting mean for temporary checkpointing and error-recovery.
Coupled applications that only act as either a data sink or a source for the main application are also major beneficiaries of burst buffers.
A prominent example in HPC are in situ visualizations copying on demand snapshots~\cite{libSim,Catalyst} or accessing the primary observables directly~\cite{ISAAC,IndeX,EAVL}.

Nevertheless, with the current absolute sizes of burst-buffers it is close to impossible to keep data between application lifetime and parallel filesystem data lifetime, simply because they cannot store a useful multiple of primary observables. As soon as a single stage in the I/O hierarchy is not drained as fast as it is filled, a backlog throughout all previous stages is inevitable even when buffers are used.


\section{I/O Measurements}

PIConGPU implements I/O for outputs and checkpoints within its plugin system.
Plugins are tightly coupled algorithms that can register within the main application for execution after selected iterations.
They share full access to primary observables (read and write) of the application.

I/O modules implemented are parallel HDF5~\cite{hdf5} and ADIOS (1.10.0)~\cite{ADIOS}.
In order to tailor domain-specific needs for particle-mesh algorithms, libSplash is used as an abstraction layer~\cite{libSplash}.
Data objects are described by the meta-data standard openPMD~\cite{openPMD100} in human- and machine-readable markup, allowing for cross-application exchangeability as needed in post- and pre-processing workflows.


\subsection{Preparation of PIConGPU Primary Observables for I/O}

In preparation of GPU device data for I/O libraries, PIConGPU field data are copied from device to host via CUDA 3D memory copies while plasma particle attributes stored in tiled data structures are copied via the mallocMC~\cite{mallocMC} heap manager.
Subsequently, scalar particle attributes are concatenated in preparation for efficient parallel I/O in a parallelized manner using OpenMP.
The single GPU data size needed for saving a complete system state is typically $S=4$~GiByte (assuming $\nicefrac{2}{3^\text{rd}}$ of device global memory for primary observables).
The overall time for preparing these 4\,GiByte of data for one GPU is typically $t_\text{prep} = 1$~s on the systems considered in this publication.

\subsection{I/O Performance in a Realistic Production Scenario}

Measurements of the I/O performance are based on one of the default benchmarks implemented in PIConGPU, a simulation of the relativistic Kelvin-Helm\-holtz Instability~\cite{PIConGPU2013,AlvesKHI}.
Starting from two spatially homogeneous, counterpropagating neutral plasma streams, a shear flow instability develops.
This scenario shows good load-balancing due to nearly homogeneous data distribution across all GPUs with data size per output and GPU of $S=4$~GiByte.
We thus assume in our following analysis for sake of simplicity that indeed each node has the same output size, the same bandwidth and I/O operations have the same impact on all $N$ nodes of a system.
\begin{table}[h]
  \caption[PIConGPU I/O benchmark systems]{PIConGPU I/O benchmark systems, both commissioned in 2012/13: relevant system characteristics and single node average filesystem throughput $T_\text{FS}$, defined as the design parallel bandwidth $B_\text{parallel}$ divided by $N$ nodes}
  \centering
  \begin{tabular}{r|l|l}
    ~ & Titan & Hypnos (queue: `k20')\\
    \hline
    GPUs / node & $1\times$ K20x & $4\times$ K20m\\
    CPUs / node & $1\times$ AMD Opteron 6274 & $2\times$ Intel Xeon E5-2609\\
    CPU-cores / GPU & 16 (8\,FP) & 2\\
    GPU / CPU Flop/s (DP) & 9.3\,:\,1 & 7.6\,:\,1\\
    file system & Spider/Lustre & GPFS\\
    $B_\text{parallel} = T_\text{FS}$ * N [GiByte/s] & 1000 & 20\\
    $T_\text{FS}$ [GiByte/s] & 0.055 & 1.25\\
    CPU $T_\text{memcpy}$ [GiByte/s] & 6.0 & 6.1\\
    maximum number of nodes $N_\text{max}$ & 18000 & 16\\
    \hline
  \end{tabular}
  \label{tab:benchmarkSystems}
\end{table}

Our benchmark systems are Titan (\gls{ornl}) and the K20 queue of Hypnos (HZDR), see Tab. \ref{tab:benchmarkSystems}.
We choose the second system intentionally, since it has roughly the same age, similar ratio of Flops/s between CPU host an GPU device, multiple GPUs per node as in upcoming systems, even less CPU cores per GPU and an even higher single node average filesystem bandwidth compared to Titan.
All measurement input and results of the following sections are available in the supplementary materials~\cite{supplementary} and all software used is open source.

Most relevant from an application point of view is the absolute overhead $t_\text{I/O}$ in seconds caused by enabling I/O since it equals `wasted' computing time that could be otherwise spent to iterate the problem further or in higher resolution.
We define the effective parallel I/O throughput $T_\text{eff}$ in GiByte/s as
\begin{align}
    T_\text{eff} \equiv N\times\frac{S}{t_\text{with\;I/O}-t_\text{without\;I/O}}=\frac{N \times S}{t_\text{I/O}}\label{eq:effBW}
\end{align}
with the number of nodes $N$, the data size per node $S$ and the difference between execution time with I/O $t_\text{with\;I/O}$ and without I/O $t_\text{without\;I/O}$ as $t_\text{I/O}$.
Besides the (included) correction for intrinsic overheads in scaling the application, all measurements are performed as a weak scaling of PIConGPU, which is near-perfect up to the full size of Titan~\cite{PIConGPU2013}.
We average over 11 outputs within 2000 iterations with an average application iteration frequency of one Hertz.

In the following we model the I/O time per node by
\begin{align}
    t_\text{I/O}^\text{simple} \equiv t_\text{prep} + t_\text{off\;RAM} = t_\text{prep} + \frac{S}{T_\text{FS}}\label{eq:OHsimple}
\end{align}
defining $t_\text{prep}$ as the time to concatenate data into large, I/O-API compatible chunks and $t_\text{off\;RAM}\equiv\nicefrac{S}{T_\text{FS}}$ as the time to synchronously send the data off RAM.
This preparation time can potentially be lowered by reorganizing data on the accelerator, where RAM is usually in full utilization from the application alone, while asynchronous (non-blocking) writes that hide data transfer latency require large enough temporary buffers to avoid backlog (see discussion in Sect. \ref{sec:backlog}) and I/O library support. It is thus $\nicefrac{S}{T_\text{FS}}$ that will dominate overhead compared to iterations without I/O.

\begin{figure}[htbp]
  \centering
  \includegraphics[width=\textwidth]{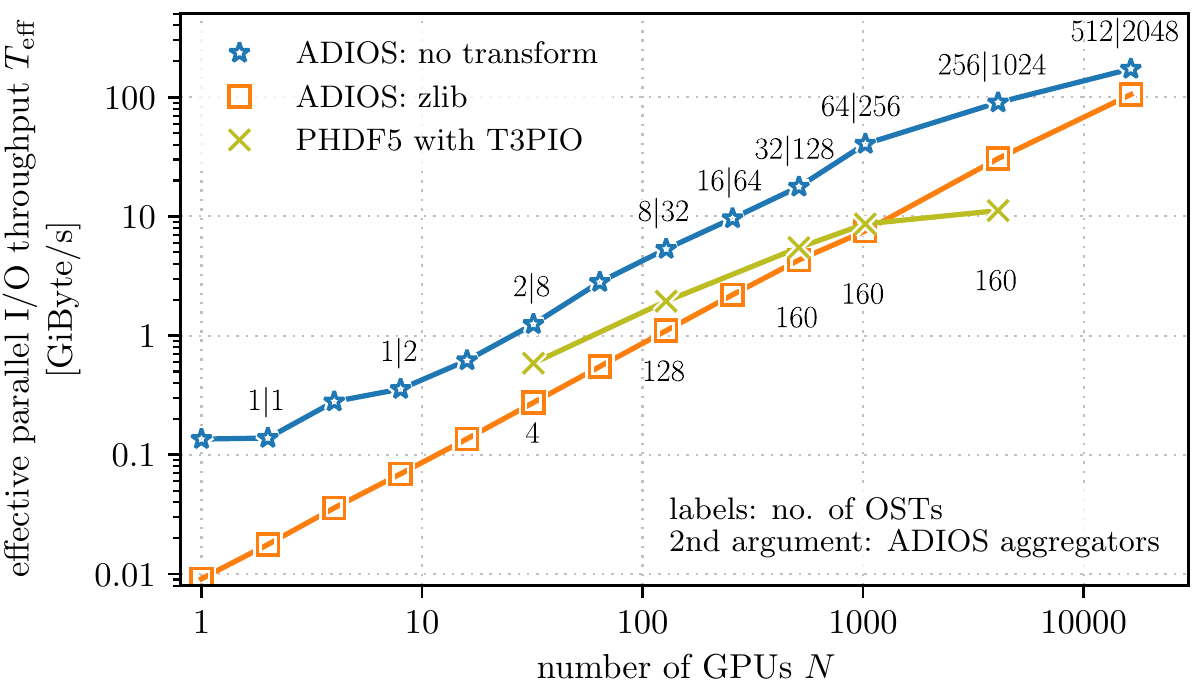}
  \caption[PIConGPU I/O Scaling on Titan]{PIConGPU I/O weak scaling on Titan from 1 to 16384 K20x GPUs (nodes).
  Zlib was only supported serially with compression mode fast.
  MPI\_Info hints for parallel HDF5 set via T3PIO (v2.3).
  For ADIOS, labels denote number of \textit{OSTs\textbar aggregators}, resulting for $N\geq 32$ in a striping of each aggregated process group over four OSTs.
  Lustre filesystem limits enforced 160 OSTs for (single-file) parallel HDF5 writes.
}
  \label{fig:bwTitanZlib}
\end{figure}

\begin{figure}[htbp]
  \centering
  \includegraphics[width=0.65\textwidth]{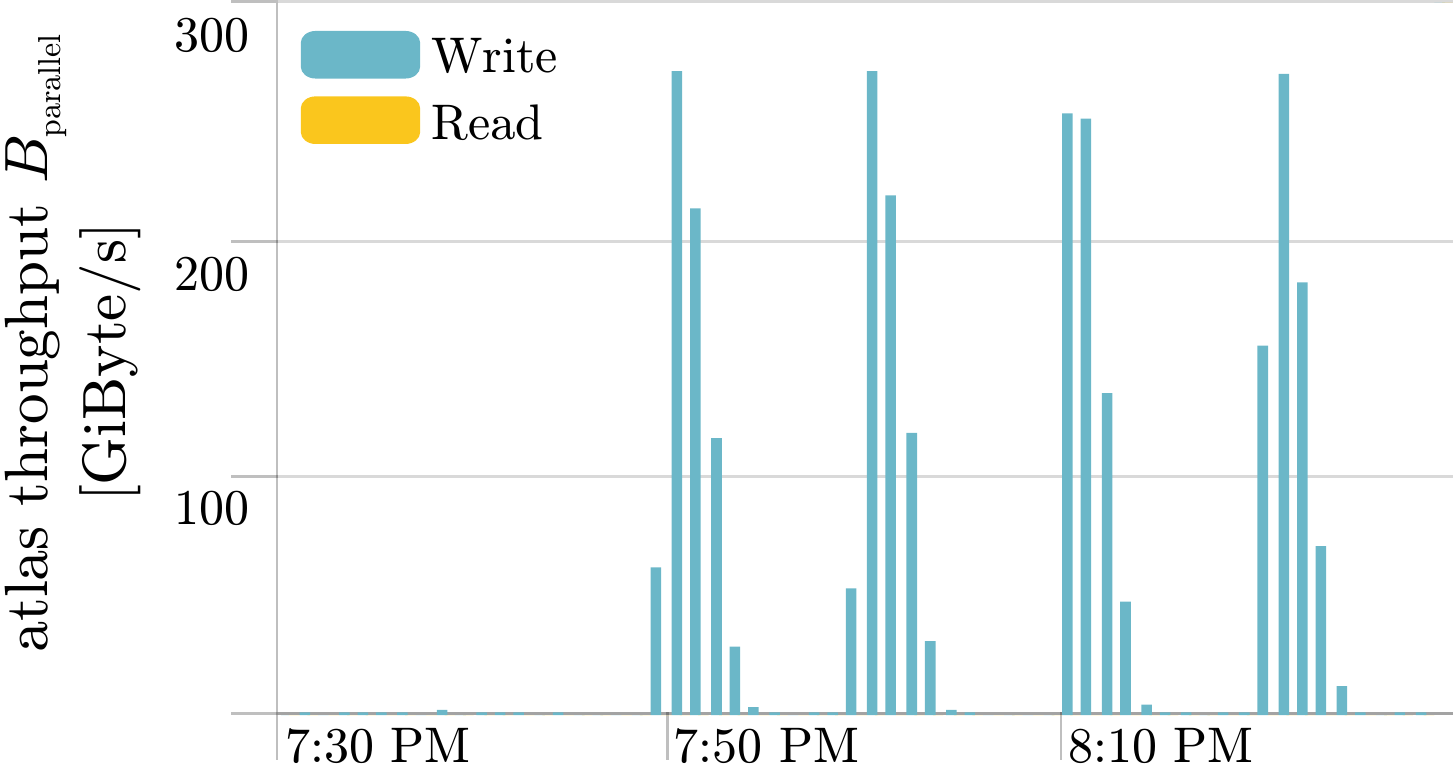}
  \caption{Actual filesystem throughput as seen by Atlas 2 (\gls{ornl}) during run no.~2489794 (Sep 23rd, 2015) on 16384 nodes according to user support (data: DDNTool, Splunk).
}
  \label{fig:bwAtlas}
\end{figure}

Figure \ref{fig:bwTitanZlib} shows the achieved effective parallel I/O throughput $T_\text{eff}$ on Titan.
We noticed HDF5 I/O overhead getting prohibitively large for production runs as its parallelism is currently limited by the number of allocatable Lustre OSTs ($\le 160$) on which one global file needs to be strided over.
After optimizing HDF5 performance with MPI-I/O and HDF5 hints, first manually via best-practices and later using T3PIO~\cite{T3PIO}, we turned down the strategy of parallel output in one global file (June 2014) and started adopting ADIOS aggregators, which enable transparently striding on subgroups of processes over a limited number of OSTs (latest benchmark: September 2015).
When using ADIOS in this manner, we were able to reach an overall application throughput close to 280\,GByte/s, see Fig.~\ref{fig:bwAtlas}.
We are not aware of substantial changes in the Atlas filesystem during this period of time, expecting both benchmarks to be comparable.

It is important to note that measuring the I/O throughput indirectly via introduced overhead masquerades the actual filesystem bandwidth $B_\text{FS}$ which is always higher than the previously defined effective parallel throughput $T_\text{eff}$ for raw, untransformed data as seen by the application.
This is very important to keep in mind as the effective parallel throughput determines the application performance in most realistic scenarios.


As mentioned in Sect. \ref{sec:iobottlenecks}, absolute I/O size during production runs quickly becomes a show-stopper.
Compressing data streams on the fly seems to suggest itself as data reduction technique, either lossless or lossy, depending on application needs.
In ADIOS, compression schemes are implemented transparently for the user as so-called data transforms.
One would not only expect a reduction in data size but also an increase in effective bandwidth since the size of the compressed data $S_\text{C}$ written to the filesystem is lowered by a compression ratio $f_\text{C}\equiv\nicefrac{S_\text{C}}{S}\leq 1$ compared to the initial size $S$.
We observed that this expectation could not be fulfilled using even the fastest compression algorithm implemented at the time in ADIOS, zlib, see Fig. \ref{fig:bwTitanZlib}.

We therefore expanded our model to account for the time $t_\text{reduce}\equiv\nicefrac{S}{T_\text{C}}$ it takes to reduce the data by compression or other means and copy it from an application-side buffer to an I/O library buffer.
Up to now, data transforms in ADIOS are performed before starting to send the data off-node, while parallel HDF5 does not yet support data compression\footnote{An experimental development preview with compression support in parallel HDF5 was announced after our measurements in February 2017.}.
In order to account for data reduction, eq.~\eqref{eq:OHsimple} needs to be extended to add synchronous reduction overhead by
\begin{align}
    t_\text{I/O}^\text{reduce} (t_\text{reduce}) &\equiv t_\text{prep} + t_\text{reduce}                                  + f_\text{C}\times t_\text{off\;RAM}\nonumber\\
                                               ~ &= t_\text{prep} + \frac{S}{T_\text{C}}                                  + \frac{f_\text{C} \times S}{T_\text{FS}}\nonumber\\
                                               ~ &= t_\text{prep} + \frac{S}{\mathcal{T}_\text{C} \times T_\text{memcpy}} + \frac{f_\text{C} \times S}{\mathcal{T}_\text{FS} \times T_\text{memcpy}}\label{eq:OH}\\
    f_\text{C}   &\equiv \frac{S_\text{C}}{S}\quad
    \mathcal{T}_\text{C} \equiv \frac{T_\text{C}}{T_\text{memcpy}}\quad
    \mathcal{T}_\text{FS} \equiv \frac{T_\text{FS}}{T_\text{memcpy}} .\nonumber
\end{align}
$\mathcal{T}_\text{C}$ and $\mathcal{T}_\text{FS}$ characterize throughput for compression and filesystem writes, respectively, normalized to in-node memory copy throughput $T_\text{memcpy}$.
We acknowledge that $t_\text{reduce} + f_\text{C}\times t_\text{off\;RAM}$ could in principle be lowered by copying the data to an I/O stage immediately and performing compression there, again within the limits of the discussion in Sect. \ref{sec:backlog}.

Consequently, for a given normalized per-node filesystem throughput $\mathcal{T}_\text{FS}$ any data reduction algorithm C needs to fulfill the relation
\begin{align}
    \frac{\mathcal{T}_\text{C} \times (1 - f_\text{C})}{1 - \mathcal{T}_\text{C}} > \mathcal{T}_\text{FS}\label{eq:breakEven}
\end{align}
in order to not only reduce data size by $f_\text{C}$ but also perceived write time.
This inequality arises from eq.~\eqref{eq:OH} assuming a reduce operation that is as fast as possible by setting the second term of the sum $t_\text{reduce}\equiv t_\text{memcpy}$ and thus comparing $t_\text{I/O}^\text{reduce} (t_\text{reduce}) < t_\text{I/O}^\text{reduce} (t_\text{memcpy})$.
The left-hand side of eq.~\eqref{eq:breakEven}, which we call the \textit{break-even threshold} for a given data transform algorithm and single (parallel) I/O stage, is discussed in greater detail in the following section.

\begin{figure}[htbp]
  \centering
  \includegraphics[width=\textwidth]{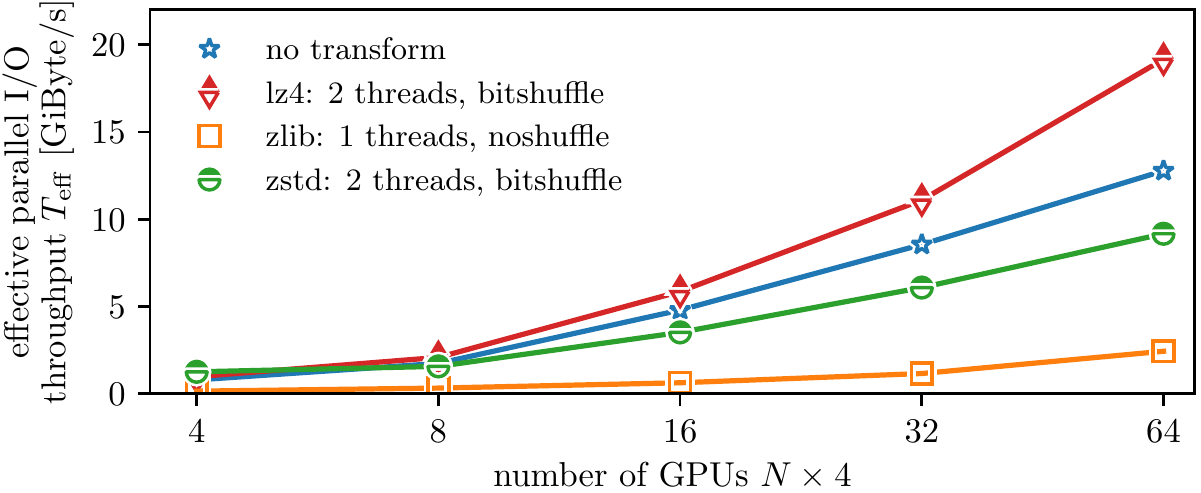}
  \caption{Weak scaling of PIConGPU with implemented I/O methods on Hypnos from 4 to 64 K20m GPUs (16 nodes).
    In contrast to Titan and Summit nodes, on Hypnos only two physical CPU cores are available per GPU, resulting in I/O performance with zlib and zstd~\cite{zstd} below the untransformed output.
}
  \label{fig:bwHypnosZlib}
\end{figure}
In order to confirm this observation, we measured I/O performance on the K20 queue of the HZDR compute cluster Hypnos, see Fig. \ref{fig:bwHypnosZlib} (data points `no transform' and `zlib').
Following eq.~\eqref{eq:breakEven} it should be even harder for a compression algorithm, lossless or lossy, to fulfill the requirements for break-even on Hypnos.
Therefore, an improvement in the latter case will be automatically favorable for Titan or a Summit-like system.

\subsection{Measurement of Compression Performance}

\begin{figure}[ht]
  \centering
  \includegraphics[width=\textwidth]{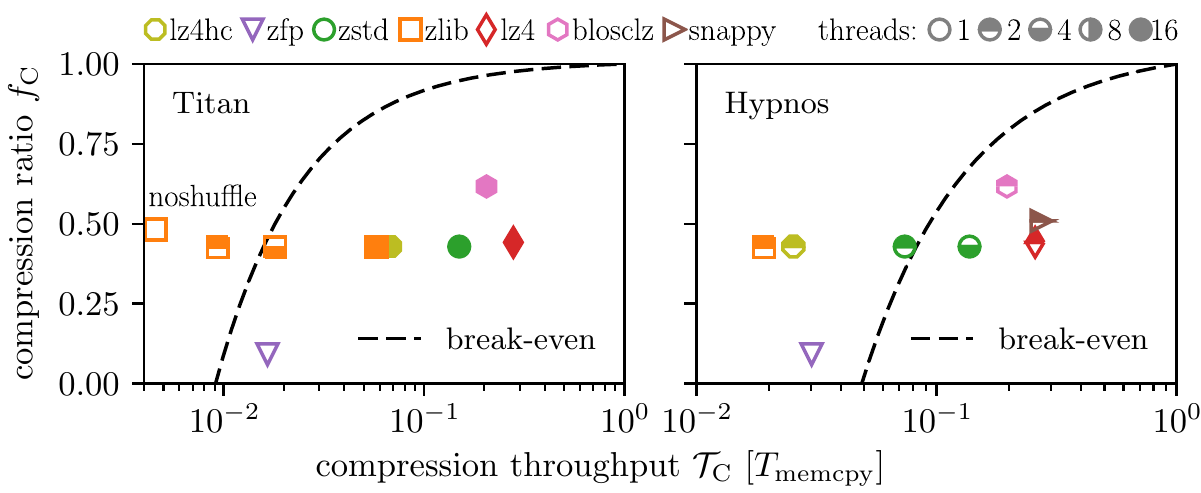}
  \caption[Compression Performance]{Compression throughput $\mathcal{T}_\text{C}$ and ratio $f_\text{C}$ measured on PIConGPU particle data (32\,Bit floating point and integers).
  Lower $f_\text{C}$ and higher $\mathcal{T}_\text{C}$ is better.
  All operations performed on contiguous, aligned, none-page-locked memory.
  The blosc~\cite{blosc} compression level is 1 (fast).
  From available pre-conditioners (none, shuffle, bitshuffle), the latter is shown due to the observed positive influence on $f_\text{C}$ with small impact on $\mathcal{T}_\text{C}$ for floating point data which otherwise could not be compressed with LZ4 (v1.7.5) and snappy~\cite{snappy}.
  Zfp (v0.5.1) was used in fixed-precision mode with three uncompressed bits per scalar~\cite{zfp}.
}
  \label{fig:compressionPerformance}
\end{figure}

In the interest of exploring feasible compression methods for PIConGPU data, we performed ex situ benchmarks on generated data.
Visualized in Fig.~\ref{fig:compressionPerformance}, such a measurement directly allows a prediction for individual systems and user data when comparing to our model, eq.~\ref{eq:breakEven}.

PIConGPU currently only utilizes one host thread per GPU, so we decided to implement and explore compression throughput for blosc as an example for a multi-threaded algorithm and compare it to other, previously implemented compression algorithms.
Blosc provides several bitshuffle pre-conditioners, which we found of great importance for floating-point compression performance in agreement with recent studies~\cite{FPcrush}.
Further benchmarks with four threads on Hypnos' K20 queue, limited to two host threads per GPU without oversubscription, indicated that on Hypnos application throughput would benefit from more physical CPU cores per GPU since the recent filesytem upgrade to GPFS.

\section{Analysis}

Fully accelerator driven applications can use 'the last 10\,\% of system performance' on the host side in order to trade compute performance for I/O latency.
The Titan system provides up to 16 physical CPU cores per GPU and Summit is expected to allow for an order of magnitude higher parallelization on the host.
This section explores the limits to data reduction methods in terms of data reduction ratio and throughput for an individual I/O stage independently of the method of data reduction and only exemplified for compression methods.


\subsection{Overhead of Compression in Parallel I/O}

From eq.~\eqref{eq:OH} the relative I/O performance ratio $\Gamma$ when using data reduction instead of direct pass-through in an I/O stage follows as:
\begin{align}
    \Gamma&\equiv\frac{t_\text{I/O}^\text{reduce} (t_\text{reduce})}{t_\text{I/O}^\text{reduce} (t_\text{memcpy})} = \frac{C_\text{prep} + \frac{f_\text{C}}{\mathcal{T}_\text{FS}} + \mathcal{T}_\text{C}^{-1}}{C_\text{prep} + \mathcal{T}_\text{FS}^{-1} + 1}\label{eq:OHrel}\\
    C_\text{prep} &\equiv \frac{t_\text{prep}}{S}\times T_\text{memcpy}\nonumber
\end{align}
where we assume that the time for reducing the data $t_\text{reduce}\geq t_\text{memcpy}$ at minimum is as long as for copying data from node RAM to I/O buffer. It is clear that in terms of I/O throughput reduction algorithms are beneficial if $\Gamma<1$ compared to I/O without reduction.
Cases of $\Gamma\geq 1$ and $f_\text{C}<1$ can still be relevant in case of limited disk space. Note, that decreasing $C_\text{prep}$ would increase the gradient of $\Gamma$, but not affect the position $\mathcal{T}_\text{C}$ for which we expect break-even.

\begin{figure}[ht]
  \centering
  \includegraphics[width=\textwidth]{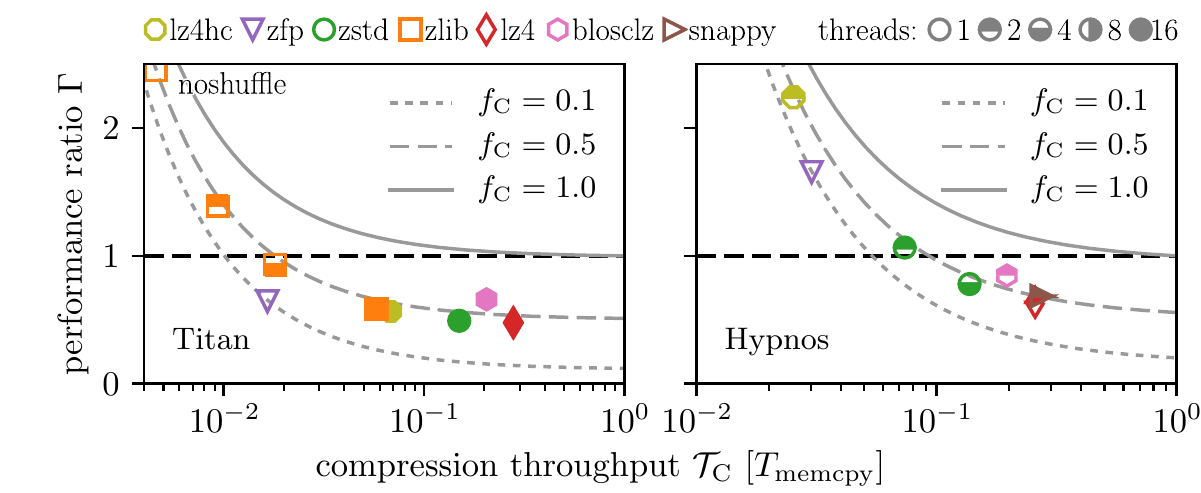}
  \caption{Visualization of eq.~\eqref{eq:OHrel} predicting the relative I/O overhead $\Gamma>1$ (gain $\Gamma\le1$) of compression during parallel I/O.
    The break-even threshold discriminates between feasible and overhead-adding compression algorithms at $\Gamma=1$ (dashed line).
    Lower $\Gamma$ and higher $\mathcal{T}_\text{C}$ is better.
    Iso-compression lines for user data are plotted for individual systems (see Tab. \ref{tab:benchmarkSystems}) and compared to measured ex situ compression performance on PIConGPU user data (see Fig. \ref{fig:compressionPerformance}).
}\label{fig:modelAnalysis}
\end{figure}
Fig.~\ref{fig:modelAnalysis} shows the effect of threaded compression, keeping the compression ratio along iso-compression lines.
Following the graph to the right, the higher the throughput of a compression algorithm the less importance it has on $\Gamma$ compared to the compression ratio $f_\text{C}$.
Thus, an important limit to $\Gamma$ is the high-throughput limit $\mathcal{T}_\text{C} \to 1$ for fast compression algorithms below the break-even threshold.
For such, the performance ratio over non-compressed I/O can barely be improved further via throughput but solely by compression ratio.

Exactly the opposite is true for any reduction algorithm with low throughput $\mathcal{T}_\text{C}$, to the left of the graph.
Above the break-even threshold (dashed line at $\Gamma=1$), data reduction quickly becomes impractical for medium to high-throughput tasks for a specific system, as the relatively wasted computing time never reaches $\Gamma\le1$ even for small $f_\text{C}$.

Following the last argument one can further derive from eq.~\eqref{eq:breakEven} with `perfect reduction' $f_\text{C} \to 0$: For \textit{any} given I/O stage with write and reduction throughput $\mathcal{T}_\text{FS}, \mathcal{T}_\text{C}$ the effective time an application spends in (synchronous) I/O can only be reduced, if the data reduction operation provides at least a throughput of
\begin{align}
    \frac{\mathcal{T}_\text{C}}{1 - \mathcal{T}_\text{C}} > \mathcal{T}_\text{FS}.\label{eq:breakEvenPerfectCompress}
\end{align}

\section{Summary and Outlook}

We implemented and benchmarked parallel I/O methods on top of state-of-the-art I/O libraries for the massively parallel, fully-manycore driven, open source \gls{pic} code PIConGPU.
We outlined performance bottlenecks for medium to high-throughput applications in general and the possibility to overcome these with general data reduction techniques such as compression.
We then derived and verified a scaling law that gives limits to expected application speed up when using data reduction schemes for medium- to high-throughput applications.
With this we were able to derive a system- and application-specific break-even threshold that allows for predicting when reducing data is benefitial in terms of I/O throughput compared to I/O without reduction.

\subsection{Compression Algorithms}

For the special case of compression algorithms, future designs to soften I/O bottlenecks first and foremost need to improve throughput for floating point data.
Even for a relatively large gap between local memory and filesystem throughput as on the current Titan system, many single-threaded compression algorithms that are still in use today do not fulfill the break-even threshold in eq.~\eqref{eq:breakEven}.

Existing high-throughput compression algorithms would benefit from research improving the compression ratio $f_\text{C}$ instead of throughput $\mathcal{T}_\text{C}$~\cite{FPcrush,Tao2017}.
This case is of importance since, due to high entropy in HPC applications' primary observables (e.g. floating point), only lossy compression algorithms are likely to bridge the upcoming throughput gaps between node-local high-bandwidth memory and storage accessible longer than application lifetime.

For ADIOS we proposed, implemented and benchmarked for the first time host-side multi-threaded transform methods as a feasible step to reach the break-even threshold.
With that, we successfully traded unused compute performance within a heterogeneous application for overall I/O performance.

\subsection{I/O Libraries}

Burst-buffers are identified as enablers to reduce blocking time of the application caused by synchronous transformations within I/O libraries, but are vulnerable to backlog.
Nonetheless, burst-buffers alone cannot cover the gap that will arise between expected I/O on system today and in the future.
Further applications of burst-buffers are coupled multi-scale simulations, in situ processing and checkpointing and not in the scope of this paper.

Nevertheless, for both explorative-qualitative and medium- to high-through\-put quantitative studies I/O libraries need to act now to provide transparent and easily programmable means for multi-stage I/O.
For any practical application, the first I/O stage should immediately start with a maximum-throughput memcopy from user RAM to I/O buffer, ideally asynchronously, while later stages need to follow fully asynchronously.
Copied memory (in unutilized RAM or burst-buffers) will need several off-node user-programmable transformations which are finally staged transparently through a subsequent non-blocking data reduction (compression) pipeline.
In each I/O stage, the break-even threshold derived in this paper needs to be fulfilled or backlog will occur for successive outputs and the overall application will be throttled by that specific bottleneck.
With deeper memory hierarchies, user-programmability of stages will be a human bottleneck and needs to be addressed with easy and fast turnaround APIs to design application- and study-specific stages, e.g. via Python/Numba.

In conclusion, introducing data reduction for I/O will be necessary because of limited medium to long term storage size expected for future systems. Our analysis and measurements show that even today one should however not expect I/O performance gains when using reduction. Parallelization of reduction algorithms is one way to gain overall I/O performance but requires compute resources in addition to those used by the application. Even for fully GPU accelerated applications one should not assume resources to be `free' for I/O and analysis tasks, since loosely coupled application workflows and models that depend heavily on hardly-parallelizable aspects such as atomic data lookups will in the future be more widespread and compete for the exact same resources.

\bibliography{citations}

\end{document}